\documentclass[twocolumn,showpacs,amsmath,prd]{revtex4}
\usepackage[dvips]{epsfig}

\begin{document}

\title{Constraints on the Extremely High-Energy Cosmic Ray Accelerators
from Classical Electrodynamics}

\author{F.A. Aharonian$^1$, A.A. Belyanin$^{2,3}$, E.V. Derishev$^{1,3}$, V.V.
Kocharovsky$^{2,3}$, Vl.V. Kocharovsky$^3$}

\affiliation{
$^1$MPI f\"{u}r Kernphysik, Saupfercheckweg 1, D-69117 Heidelberg, Germany\\
$^2$Dept. of Physics, Texas A\&M University, College Station,
TX 77843-4242, USA\\
$^3$Institute of Applied Physics RAS,
46 Ulyanov st., 603950 Nizhny Novgorod, Russia}

\begin{abstract}

We find the general requirements, set by classical electrodynamics, to
the sources of extremely high-energy cosmic rays (EHECRs).
It is shown that the parameters of EHECR accelerators are strongly
limited not only by the particle confinement in large-scale magnetic field
or by the difference in electric potentials (generalized Hillas criterion),
but also by the synchrotron radiation, the electro-bremsstrahlung, or the
curvature radiation of accelerated particles.
Optimization of these requirements in terms of accelerator's size and magnetic
field strength results in the ultimate lower limit to the overall source
energy, which scales as the fifth power of attainable particle energy. Hard
$\gamma$-rays accompanying generation of EHECRs can be used as a probe for
potential acceleration sites. We apply the results to several populations of
astrophysical objects -- potential EHECR sources -- and discuss their ability
to accelerate protons to $10^{20}$~eV and beyond. A possibility to gain from
ultrarelativistic bulk flows is emphasized, with Active Galactic Nuclei and 
Gamma-Ray Bursts being the examples.

\end{abstract}

\pacs{98.70.Sa, 95.30.Gv}

\maketitle

\section{Introduction}

The existence of EHECRs with energies
above $10^{20}$~eV is now a well-established fact (see, e.g. \cite{rev1} and
\cite{rev2} for a review). Nearly perfect isotropy of the highest energy
cosmic rays points to their likely extragalactic origin. On the other 
hand, the sources must be located within $100$~Mpc: the propagation distance
for EHECRs of energy greater than $5 \times 10^{19}$~eV is limited due to
their interaction with the microwave background radiation \cite{Gr66,ZK},
unless there are some yet unknown type(s) of particles that do not undergo
such an interaction.

The presence of non-thermal particles is a usual
feature of any plasma system. However, finding a way to accelerate these
particles to  $10^{20}$~eV is a serious theoretical challenge.
Even if an electrodynamical system used to accelerate charged particles
is sufficiently long-lived, the attainable energy is limited since the
particles can only stay within the acceleration region until their gyroradius
exceeds the size of the region. (This is known as Hillas criterion
\cite{H84}.) Various types of model- and source-dependent radiative and
collisional losses decrease the limiting particle energy furthermore. But even
if one finds a way to circumvent the latter obstacle, there are radiative
losses caused by the same electromagnetic field that is used to accelerate
particles. For example, Greisen \cite{Gr65} considered the synchrotron losses
and showed that for very small-sized accelerators, i.e., for those having very
strong magnetic field, the synchrotron radiation drains energy so fast that
even a particle accelerated at the largest possible rate for the
magnetic-field dominated environments, $qBc$ ($q$ is the particle charge, $B$
the magnetic field strength, $c$ the velocity of light), is not able to
achieve the energy given by the Hillas criterion. However, Greisen's limit
might seem to be not a general one, as in some objects accelerated particles
move along the magnetic field lines (neutron stars are a perfect example of
that), without suffering from the synchrotron losses.

In this paper, we relax the requirements to the
acceleration sites as much as possible, assuming that
inhomogeneity of electric and magnetic fields is small enough not to cause
particle diffusion out of the acceleration site and taking into account only
the radiation mechanism which gives the lowest energy loss rate.
In fact, even if the synchrotron losses (or the electro-bremsstrahlung in the
case of electric-field dominated environments) tend to zero, i.e., for
particles moving along smooth field lines, there is still curvature radiation,
which then limits the maximal particle energy. Furthermore, we generalize the
analysis introducing acceleration efficiency parameter $\eta$, so that the
acceleration rate is given by
\begin{equation}
\dot{\varepsilon} = \eta qBc\, ,
\end{equation}
where $\eta B = E_{\rm eff}$ is the projection of an electric field on the
particle trajectory, effectively averaged as the particle moves along this
trajectory. Such a representation is useful when, e.g., diffusive shock
acceleration scenario is brought into consideration.

We derive the ultimate lower limit for the total electromagnetic energy
stored in an acceleration region and find the size and the field strength that
are optimal with respect to minimization of this energy. Then we consider a
possible gain from relativistic motion of an accelerator as a whole and
discuss what kind of additional information one gets from analysis of hard
$\gamma$-photons generated as a byproduct of cosmic ray acceleration. Finally,
we make comparison of theoretical constraints with the parameters of known
astrophysical objects.

\section{Electrodynamical limitations}

Let us consider a region of space of a size $R$ where particle acceleration
occurs. A natural assumption is that there is nothing like linear accelerator
geometry in this region, but instead the field lines of both electric and
magnetic fields are curved with the curvature radius being of the order of $R$
or less.

If the magnetic field is much stronger than the electric field, $B \gg
E$, then a test particle is accelerated following  a magnetic field line in
its trajectory. If $E \gtrsim B$ but $(\vec{E} \cdot \vec{B})
\sim EB$, then particle's trajectory depends on the radiative loss rate. High
loss rate will force the particle to follow a field line (of an electric field in this
case) since a transverse to the field line component of particle's momentum
will quickly decay due to an increase in radiative losses. In the opposite case, the
field geometry will have little effect on particle's trajectory which
therefore will make an angle of $\sim 45^{\circ}$ with field lines. The
remaining case, where both Lorentz invariants $(\vec{E} \cdot \vec{B})$ and
$| E^2 - B^2|$ are much smaller than $E^2 + B^2$, can be
reduced to one of the already mentioned situations by the appropriate Lorentz
transformation to the reference frame in which the mean (after averaging in
space and time) Poynting flux density becomes zero.

In what follows we assume that accelerator's lifetime is at least as large as
the acceleration timescale and use the most optimistic premise that the
strongest of magnetic and electric fields is regular on scales $\ll R$ and
quasistatic. (Alternating field direction implies that particles escape more
readily and that the time-averaged acceleration rate is smaller, while the
radiative losses are roughly the same.) Then, the effect of small-scale
magnetic field can be neglected if $E>B$, and in the case $B>E$ the factor
$\eta < 1$ takes into account all necessary averaging for a small-scale and/or
variable electric field, so that $E_{\rm eff} = \eta B$.

The particle can be accelerated up to the terminal energy which is the
smallest of either the
work done by accelerating force in a time it takes for a particle to escape
from the region, or the energy at which radiative losses balance the
acceleration. The balance condition requires that the radiative loss rate
$\dot{\varepsilon}_{\rm rad}$ is equal to the energy gain rate, i.e.
\begin{equation}
\label{cur}
\dot{\varepsilon}_{\rm rad} = \frac{2}{3} \gamma^4 \frac{q^2}{R^2} c
= \eta qBc
\end{equation}
in the case of curvature-radiation dominated losses \cite{Ginz} (the
particle is assumed to move along an arc of radius $R$), or
\begin{equation}
\label{sy}
\dot{\varepsilon}_{\rm rad} = \frac{2}{3} \gamma^2
\left( \frac{q^2}{mc^2} \right)^2 c \left( B_{\perp}^2 +E_{\perp}^2 \right)
= \eta qBc
\end{equation}
in the case of synchrotron- or electro-bremsstrahlung dominated
losses. Here $\gamma$ is the Lorentz factor of the particle, $m$ its mass,
$B_{\perp}$ and $E_{\perp}$ are the r.m.s. field components perpendicular to
the particle momentum.

The comparison of Eqs. (\ref{cur}) and (\ref{sy}) shows that the
curvature radiation gives a more favorable (smaller) estimate for the
radiative losses as long as the particle energy, $\varepsilon = \gamma
mc^2$, satisfies the condition
\begin{equation}
\label{cur<sy}
\varepsilon <  qR \sqrt{B_{\perp}^2 + E_{\perp}^2}.
\end{equation}
If the above condition were violated, it would mean one of the following.
Either the particle is no longer bound in the acceleration region, i.e.,
its gyro-radius $r_{\rm H} = \varepsilon/ (qB_{\perp})$ is larger than $R$
(when $r_{\rm H} \gtrsim R$, the very idea of curvature radiation becomes
meaningless) and acceleration terminates upon the
particle escape at $\varepsilon \leq qR B_{\perp}$, or, if $B_{\perp} \ll
E_{\perp}$, the particle is unbound from the very beginning
and its energy gain is limited by the difference of electric potentials across
the acceleration region, $\varepsilon \leq qR E_{\rm eff} = \eta qR B$.
This provides a generalized version of the Hillas criterion.

Hereafter we neglect numerical factors of $\sim 1$ assuming $B=B_{\perp}$
and $E = E_{\perp} = E_{\rm eff} = \eta B$. In general, one should bear in mind that
for an  inhomogeneous small-scale field the r.m.s. quantity $E^2$
must be replaced with $\langle E^2 \rangle = E^2/\delta$, where $\delta < 1$
is a duty-factor. (The same is true for magnetic field, although the
situations where $E>B$ and the magnetic field has the duty-factor $\delta \ll
1$ seem quite unrealistic.) We assume below $\delta = 1$, aiming to the
most efficient acceleration. Thus, the condition given by Eq. (\ref{cur<sy})
cannot be violated, as it satisfies the generalized Hillas criterion,
\begin{equation}
\label{Hil}
\varepsilon_{\rm max} <  qR \max\{B,E\}.
\end{equation}

Since we are interested in the most efficient
acceleration (i.e., the highest acceleration rate at smallest radiative
losses), it is  Eq. (\ref{cur}) that will be used below to put  the lower
limit on the radiative loss rate. Then, we conclude that the maximal energy of
a particle  is either determined by the balance condition given by Eq.
(\ref{cur}),
\begin{equation}
\label{max-b}
\varepsilon_{\rm max}^4 = \frac{3\, \eta B R^2}{2\, q} \left( mc^2\right)^4,
\end{equation}
or is limited by the Eq. (\ref{Hil}), whichever gives the smaller value.

The overall energy of an electromagnetic field in a spherical region of radius
$R$, capable of accelerating particles up to $\varepsilon_{\rm max}$, is
$W_{\rm em} = R^3 (B^2 + E^2)/6$. According to Eqs. (\ref{max-b}) and
(\ref{Hil}), respectively, it is limited as follows:
\begin{subequations}
\label{min-en}
\begin{equation}
W_{\rm em} > \frac{2}{27}  \frac{q^2}{R}
\left( \frac{\varepsilon_{\rm max}}{mc^2} \right)^8
\frac{1+\eta^2}{\eta^2}
\end{equation}
and
\begin{equation}
W_{\rm em} > \frac{R}{6} \left( \frac{\varepsilon_{\rm max}}{q} \right)^2\, ,
\end{equation}
\end{subequations}

Our interest is to lower the energy
requirements taking $R$ as large as possible in Eq. (\ref{min-en}a) and as
small as possible in Eq. (\ref{min-en}b). Doing this, we derive the optimal
size
\begin{equation}
\label{opt-size}
R^{\rm (opt)} \simeq \frac{2}{3} \frac{\sqrt{1+\eta^2}}{\eta}
\frac{q^2 \varepsilon_{\rm max}^3}{\left( mc^2 \right)^4}\, ,
\end{equation}
such that both conditions become quantitatively the same.
This finally gives the optimal (the minimum possible)
estimate for the amount of electromagnetic energy stored in the acceleration
region:
\begin{equation}
\label{opt-en}
W_{\rm em}^{\rm (opt)} \simeq \frac{1}{9} \frac{\sqrt{1+\eta^2}}{\eta}
\frac{\varepsilon_{\rm max}^5}{\left( mc^2 \right)^4}\, .
\end{equation}
The corresponding optimal strength of the magnetic field is
\begin{equation}
\label{opt-fi}
B^{\rm (opt)} \simeq \frac{3}{2} \frac{\eta}{1+\eta^2}
\frac{\left( mc^2 \right)^4}{q^3 \varepsilon_{\rm max}^2}\, ,
\end{equation}
and $E^{\rm (opt)} = \eta B^{\rm (opt)}$.

The energy requirements given by Eqs. (\ref{min-en}) as
functions of accelerator size are presented in Fig.~1. The required
accelerator energy increases as the region size deviates from the optimal one.
A smaller region would accelerate particles to the energy $\varepsilon_{\rm
max}$ only if the energy budget is increased by $R^{\rm (opt)}/R$ times; in a
larger region, the energy budget has to be increased by $R/R^{\rm (opt)}$
times.

\begin{figure}
\label{plot}
\scalebox{0.30}{\includegraphics{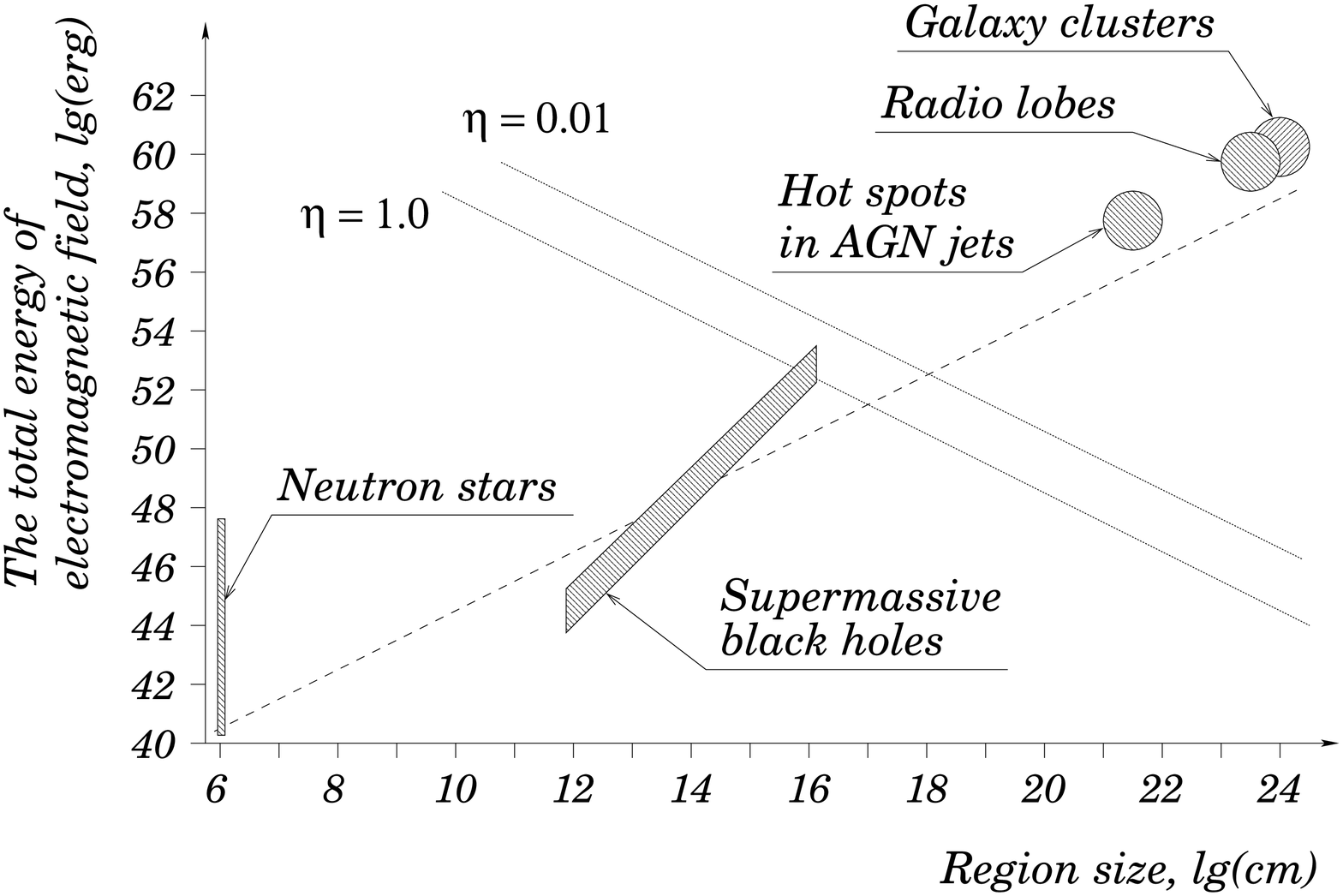}}
\caption{
The energy requirements for protons of energy $10^{20}$~eV. Eq.
(\ref{min-en}a), dotted lines (for acceleration efficiencies $\eta =1$ and
$\eta =10^{-2}$), and Eq. (\ref{min-en}b), dashed line, as functions of the
size of an acceleration region. }
\end{figure}

To have an idea of typical numbers, let us make the estimates for a proton
of energy $\varepsilon_{\rm max} = 10^{20}$~eV. In this case, the total
electromagnetic energy in the acceleration region should exceed
$W_{\rm em}^{\rm (opt)} \simeq 3 \times 10^{51}$~erg; the optimal size
$R^{\rm (opt)}$ is of the order of $10^{17}$~cm and the optimal magnetic
field is of $\sim 3$~G. When making these estimates we assumed $\eta =1$ to
lower the energy requirements. Taking $\eta$ still larger does not reduce them
too much, by a factor of 2 at most.

All the results obtained are valid also for an accelerator that moves
as a whole with the Lorentz factor $\Gamma \gg 1$, if the quantities are
measured in the comoving frame. However, it is more convenient to
present the results in a form where $W_{\rm em}^{\rm (opt)}$ and
$\varepsilon_{\rm max}$ are measured in the laboratory frame, while $R^{\rm
(opt)}$, $B^{\rm (opt)}$ and $E^{\rm (opt)}$ are measured in the comoving
frame. Then
\begin{subequations}
\label{rl-mot}
\begin{equation}
W_{\rm em}^{\rm (opt)} \simeq \frac{1}{9\, \Gamma^4}
\frac{\sqrt{1+\eta^2}}{\eta} \frac{\varepsilon_{\rm max}^5}{\left( mc^2
\right)^4}\, ,
\end{equation}
\begin{equation}
R^{\prime \rm (opt)} \simeq \frac{2}{3\, \Gamma^3}
\frac{\sqrt{1+\eta^2}}{\eta} \frac{q^2 \varepsilon_{\rm max}^3}{\left( mc^2
\right)^4}\, ,
\end{equation}
\begin{equation}
B^{\prime \rm (opt)} \simeq \frac{3\, \Gamma^2}{2} \frac{\eta}{1+\eta^2}
\frac{\left( mc^2 \right)^4}{q^3 \varepsilon_{\rm max}^2}\, ,
\end{equation}
\end{subequations}
where the primed quantities are measured in the comoving frame.
A comparison of energy requirements given by Eqs. (\ref{min-en}), generalized
for the case $\Gamma \gg 1$, for accelerators with different bulk Lorentz
factors is presented in Fig.~2.

\begin{figure}
\label{plot1}
\scalebox{0.30}{\includegraphics{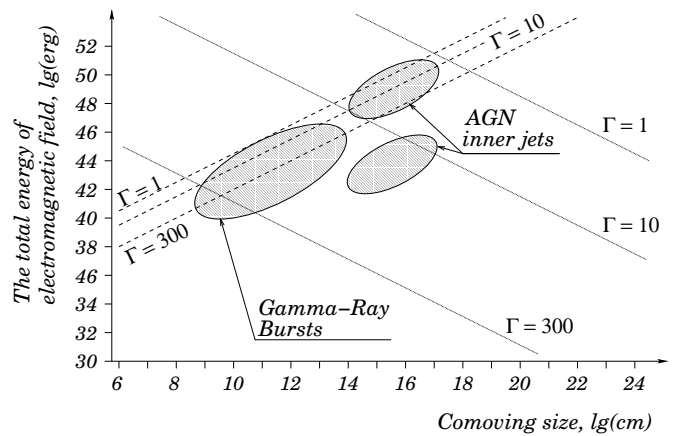}}
\caption{
The energy requirements for protons of energy $10^{20}$~eV. Eq.
(\ref{min-en}a), dotted lines (for accelerator at rest and for bulk Lorentz
factors $\Gamma =10$ and $\Gamma =300$), and Eq. (\ref{min-en}b), dashed
lines, as functions of the comoving size of an acceleration region.
For AGN inner jets, the upper zone corresponds to hadronic models of
high-energy $\gamma$-ray emission, and lower zone -- to leptonic models.}
\end{figure}

Apparently, the acceleration of EHECRs in
ultrarelativistic bulk flows has an advantage of bringing the optimal source
size to a more "comfortable" (for short-lived phenomena like Gamma-Ray Bursts)
range and reducing the energy requirements to the source. However, for
wind-like relativistic flows the actual requirements are geometry-dependent.
Indeed, for the causality reasons, the acceleration region does not occupy the
whole sphere of radius $R$, but rather extends to a distance $R^{\prime} =
R/\Gamma$ transverse to the radius and to a distance $R^{\prime}/\Gamma$ along
it, so that the total energy within the volume of radius $R$ is of $\sim
\Gamma^4 W_{\rm em}$ for a wind with a broad beam pattern. Since the energy
stored in the acceleration region, $W_{\rm em}$, is multiplied by $\Gamma^4$
in this case, one can only gain from a more favorable ratio
$R^{\prime}/R^{\prime \rm (opt)}$ (cf. Eq. (\ref{rl-mot}a)). A jet geometry is
more favorable: with the jet opening angle $\simeq 1/\Gamma$ the energy
requirements can be reduced by a factor as large as $\Gamma^2$, with respect
to the non-relativistic accelerator. To implement the advantage of
relativistic bulk motion to the full extent, i.e., to get a gain by a factor
of $\Gamma^4$, one needs a transient jet source which lasts for a time $\simeq
R/(\Gamma^2 c)$.

\section{Constraints due to radiation from acceleration sites}

Before moving to an overview of different source populations, let us
discuss some implications regarding the acceleration
timescale and the electromagnetic radiation accompanying the acceleration
process.

When $\eta \gtrsim 1$, a particle is accelerated up to
$\varepsilon_{\rm max}$ in a time $t_{\rm acc} \lesssim R/c$, which is less than or
comparable to the dynamical timescale of an accelerator. However, at low
efficiency $\eta \ll 1$, the acceleration may take a longer time, i.e.
$\eta^{-1} (R/c)$, if accelerator's size is larger than $R^{\rm (opt)}$, or
$\eta^{-1} (R/R^{\rm (opt)}) (R/c)$ otherwise. Relativistic
motion of an  accelerator usually implies relativistic expansion. Consequently, the lifetime is
of order $ R/c$, so that small $\eta$ strongly disfavors such objects unless
their size is $\lesssim \eta R^{\rm (opt)}$, that would allow a particle to attain
the energy $\varepsilon_{\rm max}$.

The acceleration of particles goes side by side with the emission of hard
electromagnetic radiation. The lower limit on the energy lost to this
radiation is given by the product of $t_{\rm acc}$ and the radiative loss rate
(Eq. (\ref{cur}), where it is convenient to substitute $\gamma^3$ from Eq.
(\ref{opt-size})). However, if $t_{\rm acc} < R/c$ one must integrate the loss
rate over a time of order $R/c$, since a particle cannot leave the
acceleration region in a shorter time. As a result of particle's acceleration,
the following fraction of its terminal energy $\varepsilon_{\rm max}$ is lost
for $\gamma$-radiation
\begin{equation}
\label{Eloss}
\frac{E_{\rm rad}}{\varepsilon_{\rm max}} \gtrsim \left\{
\begin{array}{lll}
 R^{\rm (opt)}/R
& \mbox{if} &  R>R^{\rm (opt)}\\
1 & \mbox{if} &  R^{\rm (opt)} >R> \eta R^{\rm (opt)}\\
\eta R^{\rm (opt)}/\sqrt{1+\eta^2} R
& \mbox{if} &  R< \eta R^{\rm (opt)}.
\end{array}
\right.
\end{equation}
Therefore, acceleration in small objects ($R< R^{\rm (opt)}$) is always
limited by radiative losses.

The spectral energy distribution of the
accompanying radiation has a maximum (in the source frame) around
\begin{equation}
\label{Erad}
\varepsilon_{\gamma} \sim \hbar
\frac{\gamma^3 c}{R} \sim \eta \frac{R^{\rm (opt)}}{R} \frac{mc^2}{\alpha
Z^2},
\end{equation}
provided the accelerator size is less than $R^{\rm
(opt)}$; $\alpha$ is the fine-structure constant and $Z$ the
ion charge in units of the elementary charge. The energy of curvature
radiation photons given by Eq. (\ref{Erad}) is related to the energy of
synchrotron photons as $\varepsilon_{\gamma} = (r_{\rm H}/R) \varepsilon^{\rm
sy}_{\gamma}$.

The ratio of the luminosity in
accompanying radiation to the luminosity in accelerated particles,
$L_{\gamma}/L_{\rm acc}$, as well as the energy of individual photons, is
reduced if the acceleration time is limited to $R/c$, like in the sources with
relativistic bulk motion (particle diffusion may lead to analogous limit), but
this is at expense of reducing $\varepsilon_{\rm max}$. Since the radiative
loss rate (Eq. (\ref{cur})) and the energy of individual photons (Eq.
(\ref{Erad})) scale as $\gamma^4$ and $\gamma^3$, while particle's Lorentz
factor increases linearly with time, the corresponding factors for
$L_{\gamma}/L_{\rm acc}$, $\varepsilon_{\gamma}$ and $\varepsilon_{\rm max}$
are: $(\eta R^{\rm (opt)}/R)^4$, $(\eta R^{\rm (opt)}/R)^3$, and $(\eta R^{\rm
(opt)}/R)$ respectively, in the case where $R^{\rm (opt)} > R > \eta R^{\rm
(opt)}$. Note that if the maximal energy were limited by the synchrotron
rather than curvature losses, then the acceleration up to $\varepsilon_{\rm
max}$ would not be possible at all in the objects smaller than $R^{\rm (opt)}$
if $\eta \leq 1$, as noted in \cite{Gr65}.

The check whether the sky-averaged luminosity in accompanying
electromagnetic emission exceeds the diffuse background is a powerful
selection tool for possible acceleration sites. For example, the acceleration
of protons with high efficiency, $\eta \simeq 1$, results in
$\gamma$-radiation in energy range $\gtrsim 100$~GeV (Eq. (\ref{Erad})). Such
photons undergo reprocessing into $10$~MeV - $100$~GeV energy range through
interaction with cosmic infrared and microwave backgrounds. The observed
spectral energy distribution of diffuse $\gamma$-ray background, $\sim
10^{-9}$~erg/(cm$^{2}$~s~sr)$^{-1}$ \cite{gamma-back}, is virtually constant
in the $10$~MeV - $100$~GeV spectral domain, and the observed flux of
$10^{20}$~eV cosmic rays is $\sim 2-5\times
10^{-12}$~erg/(cm$^{2}$~s~sr)$^{-1}$ \cite{rev2}, i.e., $200-500$ times less.
But the GZK effect limits the distance to which the sources of $10^{20}$~eV
cosmic rays can be observed to $\sim 30$~Mpc, while the related
$\gamma$-radiation is visible up to a redshift $z \sim 1$, i.e. the distance
$\simeq 3$~Gpc. If the accelerators are uniformly distributed in the Universe
and no other contributions to the $\gamma$-ray background are taken into
account, this implies $L_{\gamma}/L_{\rm acc} \lesssim 5$ and hence $R/(\eta
R^{\rm (opt)}) \gtrsim 1/5$. Since the spectral energy distribution of the
$\gamma$-ray background is nearly constant in $10$~MeV - $100$~GeV range, the
same limit applies for iron nuclei, although their individual
$\gamma$-luminosity is concentrated in the lower-energy spectral domain
$\gtrsim 10$~GeV (see Eq. (\ref{Erad}), which gives scaling
$\varepsilon_{\gamma} \propto A/Z^2$).

The accompanying radiation may also help in identifying acceleration sites.
The electrons are more suitable for this purpose as they can achieve
radiative-loss limited regime of acceleration in an object with much smaller
energy content. For example, there is an excess emission in $100$~MeV range
from the Crab plerion, which was interpreted as the electron synchrotron
radiation \cite{crab100}. Interestingly, the size of the plerion, $\sim
10^{17}$~cm, and its estimated magnetic field, $\sim 10^{-4}$~G, are very
close to the optimal values for electrons accelerated at the efficiency $\eta
=1$. The electrons then attain the energy $\simeq 3 \times 10^{15}$~eV and
the peak of their synchrotron spectrum is at $\simeq 100$~MeV.

So far, we did not take into account the proton interaction with
smallest-scale electromagnetic field, which results in photomeson reactions
(at wavelengths $<0.04$~cm or photon energies $>2 \times 10^{-3}$~eV for
$10^{20}$~eV proton). Such an interaction, with the
cross-section $\sigma_{p\gamma} \simeq 2\times 10^{-28}$~cm$^2$ and
inelasticity $f \simeq 0.5$, may in some cases lead to unacceptably high
energy losses for the accelerated particles. For illustration, we make the
estimate for $10^{20}$~eV protons and for the most favorable case $\eta=1$. In
order to allow accelerated protons to escape without significant losses, the
luminosity of the source at the maximum of integral photon spectrum must be
\begin{equation}
L < \pi
\frac{Rc}{f\sigma_{p\gamma}} \bar{\varepsilon} \simeq 1.5 \times 10^{44}
\left( \frac{R}{10^{17}\, \mbox{cm}} \right) \left(
\frac{\bar{\varepsilon}}{1\, \mbox{eV}} \right) \mbox{erg/s}\, ,
\end{equation}
where $\bar{\varepsilon}$ is the location of the maximum.

\section{Overview of possible acceleration sites}

{\bf Neutron stars.} Because of their size, which is far too small compared to
$R^{\rm (opt)}$, the performance of neutron stars as cosmic ray accelerators
is limited be the curvature losses (see, e.g., \cite{Ginz}). For $10^{20}$~eV
EHECRs, the energy requirements to neutron star magnetospheres are unphysical:
$\sim 10^{62}$~erg for protons and $\sim 10^{51}$~erg for iron nuclei.
Acceleration of heavier nuclei, with mass $A \sim 150$ and charge $Z \sim 50$,
to slightly smaller energy $5 \times 10^{19}$~eV is marginally possible for
rapidly rotating (i.e. $\eta \sim 1$) neutron stars with magnetic field $B
\sim 10^{15}$~G (so-called magnetars). However, the question whether such
nuclei can be efficiently transported to neutron star surface and ejected from
it remains open. In addition, the luminosity in accompanying $10 - 100$~TeV
curvature $\gamma$-radiation (which is likely to be reprocessed through
electromagnetic cascade into $1-100$~MeV range) would be 5 to 6 orders of
magnitude higher than in the produced EHECRs. Particles accelerated near the
light cylinder produce less accompanying radiation, but the energy
requirements rise in this case. Therefore, magnetospheres of neutron stars
have to be excluded from the list of EHECR sources.

A way out from the problem of catastrophic overproduction of the accompanying
radiation is acceleration of particles within ultrarelativistic pulsar winds.
However, the energy constraints, which now have to be reformulated in terms of
required Poynting-flux luminosity $L_{\rm em} \sim \Gamma^4 W_{\rm em} c/R$
for a wind with a broad beam pattern, remain extremely tough. The best case
($R^{\prime}=R^{\prime \rm (opt)}$ and $\eta=1$) for $10^{20}$~eV protons
corresponds to $\Gamma \lesssim 600$ and $L_{\rm em} \simeq 10^{45}
\Gamma^2$~erg/s. With iron nuclei the figures become $\Gamma \lesssim 60$ and
$L_{\rm em} \simeq 1.5 \cdot 10^{42} \Gamma^2$~erg/s. In any case, the
required luminosity is far beyond any of the observed values.

{\bf Black holes.} Rapidly rotating black holes embedded in a magnetic field
can generate electric field and accelerate particles (e.g.,
\cite{BHacc,BHacc1}). We make an (optimistic) estimate assuming that the 
magnetic field is roughly in equipartition with  accreted plasma, an accretion
rate corresponds to the Eddington regime, and acceleration occurs within $3\,
R_g$, where $R_g \simeq 3\times 10^{5}$~cm~$M/M_{\odot}$ is the Schwarzschild
radius of the black hole of mass $M$. The energy $W_{\rm em} \sim \pi m_p c^2
R_g^2/ \sigma_T$ is available in the acceleration region ($m_p$ is the proton
mass, and $\sigma_T \simeq 6.6 \times 10^{-24}$~cm$^{2}$ the Thomson
cross-section). This is to be compared with the required energy $(R^{\rm
(opt)}/3\, R_g) W_{\rm em}^{\rm (opt)}$, which yields
\begin{equation}
\begin{array}{l}
\displaystyle
R_g > \left( \frac{2}{81\, \pi} \sigma_T \frac{1+\eta^2}{\eta^2}
\frac{q^2 \varepsilon_{\rm max}^8}{\left( mc^2\right)^8 m_p c^2}
\right)^{1/3}\\[0.5cm]
\displaystyle
\phantom{R_g} \simeq 3 \times 10^{15} Z^{2/3} A^{-8/3}\, \mbox{cm}.
\end{array}
\end{equation}
If the accelerated particles are protons, then only super-massive black holes
with $M > 10^{10} M_{\odot}$ can meet the above requirement. On the other
hand, a lighter hole of $M > 2 \times 10^{6} M_{\odot}$ can
accelerate iron nuclei, and, since the size of such a hole is close to the
optimal one or exceeds it ($R_g \gtrsim R^{\rm (opt)} \simeq 8\times
10^{12}$~cm), this would not be at odds with the observed $\gamma$-ray
background.

{\bf Active Galactic Nuclei (AGNs).} This phenomenon is
due to ejection of a relativistic jet by a massive black hole interacting
with an accretion disk around it. Inner jets in AGNs have typically
Lorentz factors between 5 and 10. The estimate of the magnetic field depend on
the radiation model. For leptonic models which imply high-energy $\gamma$-ray
production through inverse Compton scattering, the magnetic field is of
$\sim 0.1$~G (e.g., \cite{AGNsmall}). However, in hadronic models, in
particular in the extreme proton-synchrotron model, the field strength exceeds
$100$~G \cite{AGNlarge}. Furthermore, the size of acceleration region,
$10^{14} - 10^{16}$~cm as inferred from variability timescale, is close to the
optimal value for $10^{20}$~eV protons. This eliminates
problems with $\gamma$-ray background and minimizes energy requirements that
constrain the Poynting-flux luminosity of the source in the case of a
relativistic outflow. This luminosity ($\Gamma W_{\rm em} c/R^{\prime}$) must
be higher than $\sim 10^{45}$~erg/s (or a factor of $(\theta\Gamma)^2$ more if
the opening angle of the jet, $\theta$, is larger than $\Gamma^{-1}$), which
is reasonable for a typical AGN.

{\bf Large accelerators.} There are three kinds of very large objects that
were proposed to produce EHECRs: knots (or hot spots) in large-scale AGN jets,
radio lobes (shocked regions), and galaxy clusters (e.g.,
\cite{Large,H84,Bier}). Despite their sizes are far from the optimum: $\sim
1$~kpc, $\sim 100$~kpc, and $\sim 1$~Mpc respectively, the energy of a
magnetic field is enough to bind $10^{20}$~eV protons. The limits on these
objects are better formulated in terms of the requirements to the acceleration
efficiency $\eta$, and they are likely to arise from particle diffusion. In
all cases there is about 1000-fold excess of the energy stored in magnetic
field with respect to that shown in Fig.~1. Therefore, if particles
escape on the Bohm-diffusion timescale $\sim 3\, R^2/(r_H c)$, a rather high
acceleration efficiency $\eta \gtrsim 0.03$ is necessary. Even if the
diffusion is strongly suppressed (e.g., special configuration of magnetic
field lines), there are limits on $\eta$. It must be larger than $\sim (0.3
-1) \times 10^{-2}$ in radio lobes and galaxy clusters, otherwise protons are
accelerated so slowly that they start to lose energy interacting with the
cosmic microwave background, which eventually stops acceleration. In hot
spots, with their stronger magnetic field up to $B \sim 10^{-3}$~G, $\eta$
can be as small as few~$\times 10^{-5}$, in which case their size becomes
slightly smaller than the optimal one leading to a further increase in energy
requirements.

{\bf Gamma-Ray Bursts (GRBs).} These explosion-like events have been argued to
be suitable places for acceleration of EHECRs (e.g., \cite{Wax, Viet}), owing
to their compactness (variability timescale less than a second), high
apparent ($4\, \pi$)
luminosity of the order of $10^{52}$~erg/s, and ultrarelativistic bulk flows
with the Lorentz factors $\Gamma \sim 100 - 1000$, which are the key
ingredients of any GRB model (see, e.g., \cite{GRBrev} for a review).
Acceleration may take place either in internal
shocks formed within the outflow (at a distance $10^{10}-10^{13}$~cm from the
central engine) by colliding shells of different Lorentz factors or in the
external shock (at a distance $10^{16}-10^{17}$~cm) at the interface between
ejected material and interstellar gas. The distance from which the accelerated
protons can escape is, however, limited at least by their interaction with the
intense thermal photon field left after adiabatic expansion of relativistic
fireball. The escape distance is (in the shock comoving frame)
\begin{equation}
\label{escape}
R^{\prime}_{\rm esc} \simeq 0.08 \frac{f\sigma_{p\gamma}L}{\pi \Gamma^3 c\, T}
\simeq \frac{2\times10^{17}\, \mbox{cm}}{\Gamma^3}
\frac{L}{10^{52}\, \mbox{erg/s}} \frac{3\, \mbox{MeV}}{T},
\end{equation}
where $L$ is the GRB $4\, \pi$-luminosity and $T$ the temperature at the base
of the fireball, $T \simeq 3$~MeV. The Eqs. (\ref{rl-mot}b) and (\ref{escape})
scale with the same power of $\Gamma$, so that the ratio between
$R^{\prime}_{\rm esc}$ and $R^{\prime \rm (opt)}$ depends practically only on
GRB luminosity, and the escape distance appears to be larger than the optimal
size, unless a GRB has unusually low luminosity. Therefore, radiative losses
are negligible and the maximal particle energy is $\varepsilon^{\prime}_{\rm
max} = \eta q B^{\prime} c (R^{\prime}/c)$. With the magnetic field related to
the GRB $4\, \pi$-luminosity via the parameter $\epsilon_m$, which is the
fraction of energy carried in the form of magnetic field, this gives
\begin{equation}
\label{Egrb}
\varepsilon_{\rm max}= \Gamma \varepsilon^{\prime}_{\rm max} =
\frac{\eta \epsilon_m^{1/2}}{\Gamma} q
\sqrt{\frac{2\, L}{c}} \sim \frac{2.5\times10^{23}\, \mbox{eV}}{\Gamma}.
\end{equation}
Note that the maximal energy $\varepsilon_{\rm max}$ decreases with
increasing bulk Lorentz factor $\Gamma$. This can be understood following the
same reasoning as at the end of the Sect.~II: changing the Lorentz factor
allows one to gain only from a more favorable ratio $R^{\prime \rm
(opt)}/R^{\prime}$, but for GRBs the escape distance is larger than the
optimal size, so that the increase in $\Gamma$ makes things worse.

Since explanation of some GRB
properties, such as rapid variability and the presence of emission in the
$10$~GeV domain \cite{Hurley}, requires $\Gamma > 100$, we conclude that GRBs
might be capable of accelerating protons well above $10^{20}$~eV.
However, to match the observed cosmic ray flux at least 50 per cent of the
observed GRB energy has to be converted into EHECRs.

\section{Conclusion}

Radiation losses, either due to synchrotron or curvature radiation, play
increasing role in the energy balance of accelerated particles with decreasing
size (and accordingly increasing magnetic field) of the accelerator. In
sufficiently compact objects the radiative losses limit maximal particle
energy to the values that are smaller than those determined from the Hillas
criterion. Thus, there are optimal size and magnetic field strength that
minimize the requirements to the internal energy of the accelerator. For
$10^{20}$~eV protons and acceleration efficiency $\eta = 1$, the optimal size
is $R^{\rm (opt)} \simeq 10^{17}$~cm and the optimal magnetic field is $B^{\rm
(opt)}\simeq 3$~G. This corresponds to the total electromagnetic energy in the
acceleration region $W_{\rm em}^{\rm (opt)} \simeq 3 \times 10^{51}$~erg.
Relativistic bulk motion reduces the energy requirements and the optimal
comoving size, while increasing the optimal comoving magnetic field. The
Lorentz factor $\Gamma=10$ leads to $R^{\prime \rm (opt)} \simeq 10^{14}$~cm
and $B^{\prime \rm (opt)} \simeq 300$~G, corresponding to the total
electromagnetic energy $W_{\rm em}^{\rm (opt)} \simeq 3 \times 10^{47}$~erg in
the region of size $R^{\prime \rm (opt)}$.

Efficient acceleration in optimally-sized objects leads to significant
radiative losses, which in this case occur in a regime transitional between
the synchrotron and the curvature radiation. The emission is concentrated at
photon energies of $\sim 100-300$~GeV (protons) and $\sim 50-150$~MeV
(electrons) times the bulk Lorentz factor. This suggests the use of new
generation of air-Cherenkov telescopes and GLAST, respectively, for a search
for efficient accelerators.

The derived constraints seem to firmly exclude any known population of neutron
stars as acceleration sites of EHECRs. Black holes also appear to be not able
to accelerate protons to $10^{20}$~eV except for the supermassive ones
with masses $M > 10^{10}M_{\odot}$, although acceleration of heavy nuclei
by lighter holes is still a possibility. Other astrophysical objects
considered above are able, in principle, to accelerate EHECRs,  although the
required efficiency $\eta$ is always comparable to unity. In the
standard treatment of the problem, such an efficiency is difficult to achieve
even in ultrarelativistic shocks \cite{rel-eta}. In addition, energy losses
due to photomeson reactions may be a serious problem for radiation-rich
environments of supermassive black holes and for GRB sources.

The results obtained in this paper are the unavoidable limits, set
by the laws of classical electrodynamics. Taking into
account the effects of diffusion, the energy losses via photomeson process,
the energy content of particles and waves in cosmic plasma, etc., should lead
to still stronger limits. However, they cannot, of course, constrain the
top-down models for the origin of EHECRs (e.g., \cite{top-down}), which make
use of quantum effects and non-electromagnetic interactions.

In view of the development of advanced experiments which aim at detection of
cosmic rays (CRs) above $10^{21}$~eV, it is worth noting what is the highest
energy of EHECRs still consistent with the acceleration scenario of their
origin. Plotting diagrams similar to Figs.~1 and 2, but for different values
of $\varepsilon_{\rm max}$, one can see that in GRBs and AGN inner jets
protons can in principle be accelerated up to $10^{21}$~eV. At such energies,
acceleration is likely to be radiative-loss limited in both cases. Above
$10^{21}$~eV GRBs fail because of their insufficient duration and AGN inner
jets -- because of insufficient Poynting-flux luminosity. Hot spots, radio
lobes and galaxy clusters can still work to $(3-5) \times 10^{21}$~eV under
very speculative assumption that the magnetic field is ordered on all scales
and the acceleration efficiency is $\eta \simeq 1$. In this case, acceleration
is escape-limited. At the energies $\gtrsim 10^{22}$~eV the CR primaries have
to be heavy nuclei. In all the sources listed above the heavy nuclei are
accelerated in the escape-limited regime, so that the attainable energy is
roughly $Z$ times more than for protons. However, the nuclei of such energy
are fragmented through interaction with the microwave background photons after
traveling a distance of less than $1$~Mpc, that means they must be
produced within the local group of galaxies, and GRBs is the only possibility
to do this. Although the nuclei are easily fragmented in radiation-reach
environments of GRBs, we have to conclude that formally the primaries with
energy up to $(2-3) \times 10^{22}$~eV can be produced within the framework of
acceleration scenario.

\end{document}